\documentclass[twocolumn, pra, amssymb, superscriptaddress, aps, showpacs,preprintnumbers,
amsmath,showkeys,floatfix]{revtex4}

\newcommand{\Xstate}{\ensuremath{X^2\Sigma_g^+}\,}
\newcommand{\Astate}{\ensuremath{A^2\Pi_u (\nu=2)}\,}
\newcommand{\Bstate}{\ensuremath{B^2\Sigma_u^+ (\nu=0)}\,}
\newcommand{\AstateEl}{\ensuremath{A^2\Pi_u}\,}
\newcommand{\BstateEl}{\ensuremath{B^2\Sigma_u^+}\,}
\newcommand{\Np}{\ensuremath{\text{N}_2^+}\,}

\usepackage{epstopdf}
\usepackage{graphics}
\usepackage{graphicx}
\usepackage{dcolumn}
\usepackage{bm}
\usepackage{longtable}
\usepackage{epsfig}
\usepackage{times}
\usepackage{url}
\usepackage{color}

\begin{document}
\title{Ramsey interferometry through coherent $X^2\Sigma_g^+ - A^2\Pi_u - B^2\Sigma_u^+$ coupling and population transfer in N$^+_2$ air laser}
	
\author{Yu-Hung~Kuan}
\affiliation{Department of Physics, National Central University, Taoyuan City 32001, Taiwan}
\author{Xiangxu~Mu}
\affiliation{State Key Laboratory for Mesoscopic Physics, School of Physics, Peking University, Beijing 100871, People’s Republic of China}
\author{Zhiming~Miao}
\affiliation{State Key Laboratory for Mesoscopic Physics, School of Physics, Peking University, Beijing 100871, People’s Republic of China}	
\author{Wen-Te~Liao}
\email{wente.liao@g.ncu.edu.tw}
\affiliation{Department of Physics, National Central University, Taoyuan City 32001, Taiwan}
\author{Chengyin~Wu}
\affiliation{State Key Laboratory for Mesoscopic Physics, School of Physics, Peking University, Beijing 100871, People’s Republic of China}
\author{Zheng~Li} 
\email{zheng.li@pku.edu.cn}
\affiliation{State Key Laboratory for Mesoscopic Physics, School of Physics, Peking University, Beijing 100871, People’s Republic of China}
\date{\today}
\begin{abstract}
The laser-like coherent emission at 391nm from N$_2$ gas irradiated by strong 800nm pump laser and weak 400nm seed laser is theoretically investigated. Recent experimental observations are well simulated, including temporal profile, optical gain and periodic modulation of the 391nm signal from N$_2^+$. Our calculation sheds light on the long standing controversy on whether population inversion is indispensable for the optical gain. We demonstrate the Ramsey interference fringes of the emission intensity at 391nm formed by additionally injecting another 800nm pump or 400nm seed, which are well explained by the coherent modulation of transition dipole moment and population between the $A^2\Pi_u(\nu=2)$-$\Xstate$ states as well as the \Bstate-\Xstate states. This study provides versatile possibilities for the coherent control of \Np air laser.
\end{abstract}
\maketitle
	
\begin{figure}
\vspace{-0.3cm}
\includegraphics[width=0.5\textwidth]{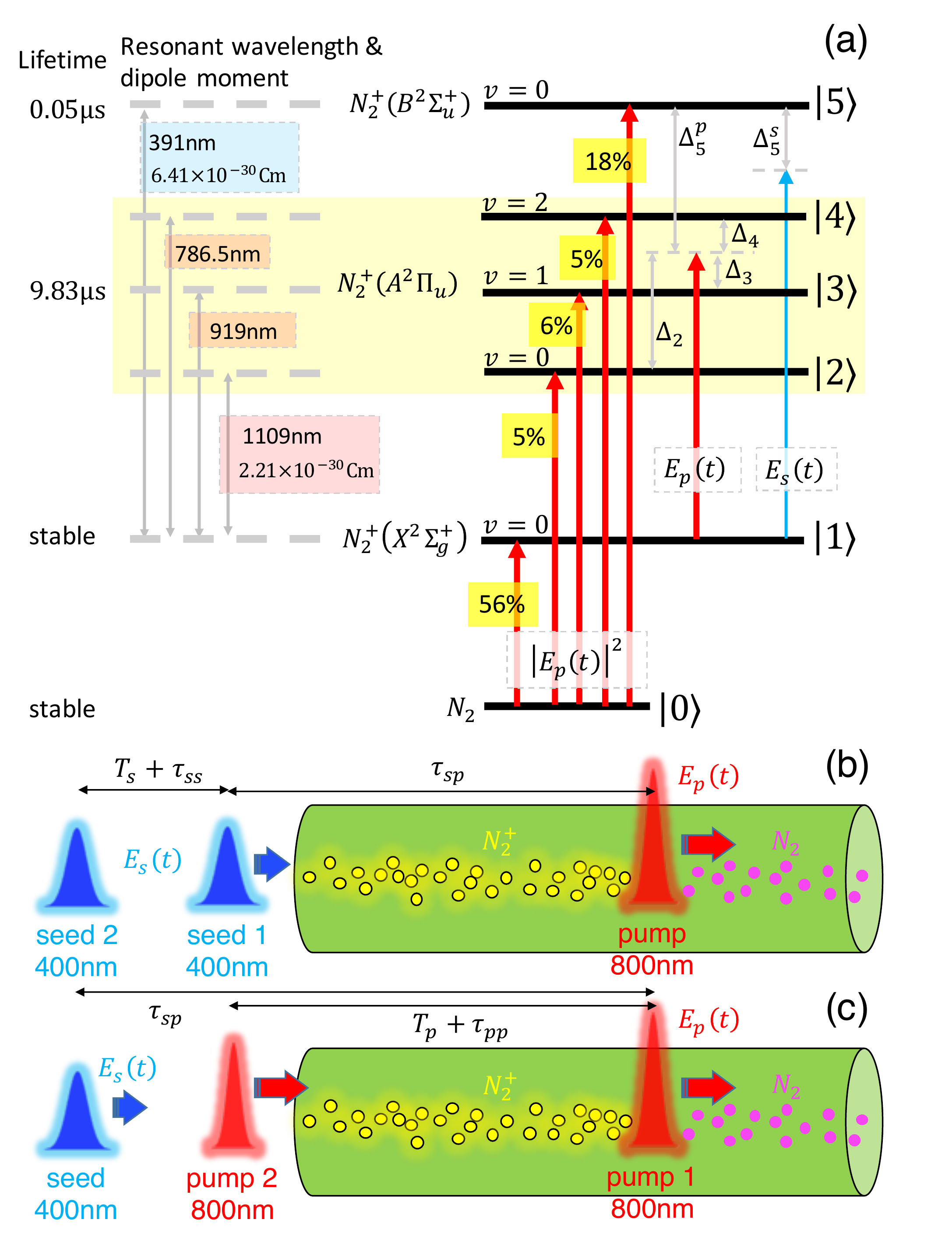}
\caption{\label{fig1}
(Color online) 
(a) N$^+_2$ level scheme. A 800nm pump pulse $E_p$ (red upward arrow) simultaneously ionizes N$_2$, with corresponding branching ratio indicated by yellow-background percentages,
and coherently drives N$^+_2$ transitions 
with detunings $\Delta_2$, $\Delta_3$, $\Delta_4$ and $\Delta^p_5$, respectively. A 400nm seed pulse $E_s$  (blue upward arrow) drives transition $\vert 1\rangle \rightarrow \vert 5\rangle$ with detuning $\Delta^s_5$ and induces emission at wavelength 391nm.
(b) double-seed (blue Gaussians) and one pump pulse (red Gaussian) scheme. Red and yellow dots depict  N$_2$ and N$^+_2$, respectively.
$\tau_{sp}$ and $T_s+\tau_{ss}$ are delay time between the pump pulse and the first seed pulse and that between two seed pulses, respectively. (c) one seed and double-pump pulses scheme. $T_p+\tau_{pp}$ is the delay time between two pump pulses.
		}
\end{figure}

A plasma filamentation will be formed when strong femtosecond laser pulses propagate in air, in which Yao et al. \cite{Yao2011} firstly reported the directional narrow-band emissions from singly ionized nitrogen molecules. This kind of mirror-free coherent emission of ambient air has potential application in remote sensing and has attracted great interest. The generation of the coherent emission around 391 nm, corresponding to the transition of \Np ($ B^2\Sigma_u^+ ,\nu''=0\ \to X^2\Sigma_g^+ , \nu=0$), has been extensively investigated\cite{Liu2015,Liu2017,Xu2015,Azarm2017,Britton2018,Zheng2019,Chen2019,Li2019,Yao2018APB,Yao2018NJP,Liu2018,AndoToshiaki2019RVaE,ZhangAn2019Cmos,ZhangAn2019Smos,Yao2016}. The underlying mechanism is much more complicated when the pump laser wavelength is around 800 nm~\cite{Yao2018APB}. The strong laser field will cause coherent coupling of the electronic, vibrational, and rotational freedom of the ionized nitrogen molecules. The coherent emission around 391 nm can be amplified by two or three orders of magnitude when injecting a seed laser with spectrum covering 391 nm. Further it is found that the amplified emission lags behind the seed for several picoseconds and lasts tens of picoseconds. 
{The amplified coherent emission is triggered in a pencil-like sample, as the macroscopic polarization is established by the seed, the effect is experimentally identified as superfluorescence~\cite{Liu2015,ZhangAn2019Smos,ZhangAn2019Cmos}.}
Even though partial evidences support the establishment of  population inversion between \Np ($ B^2\Sigma_u^+ ,\nu''=0$) and \Np ($  X^2\Sigma_g^+ , \nu=0$) by the 800 nm pump laser, it is still under hot debate whether population inversion is indispensable for the optical gain because superfluorescence could in principle be formed without population inversion. The post-ionization population redistribution mechanism (PPRM) is proposed to interpret the establishment of the population inversion\cite{Xu2015,Yao2016}, in which the population is transferred from \Np (\Xstate) to \Np (\AstateEl) by the 800 nm pump laser. The population redistribution helps to build the population inversion between \Np ($B^2\Sigma_u^+ ,\nu''=0$) and \Np (\Xstate$,\nu=0$). Very recently, the 
crucial involvement of the \Np (\AstateEl) state in the lasing process is verified in the experiments with 800nm pump field\cite{Chen2019,Li2019,AndoToshiaki2019RVaE}. Further, some periodic modulations of the 391nm coherent emission are observed in experiments by double-seed and double-pump schemes~\cite{ZhangAn2019Cmos,ZhangAn2019Smos}.

Here we provide theoretical study of effects of $A^2\Pi_u (\nu=0,1,2)$ state to the 391 nm optical gain based on the double-pump-seed scheme, and coherence effects induced by pair of seed pulses based on pump-double-seed scheme. We also focus on the population evolution induced by the 800nm pump pulse, and its consequences to the 391nm  emission.
The energy level diagram of the mixed N$_2$-\Np system is shown in Fig.~\ref{fig1}(a), where we include three vibrational states of $A^2\Pi_u$, labeled as $|2\rangle$, $|3\rangle$, $|4\rangle$. 
The \Xstate-\Astate transition is resonant with the 800nm pump and the second weak pump pulse, which has a spectral range from 770nm to 830nm.
In Fig~\ref{fig1}(b), we show the pulse sequences of the double pump-seed and the pump-double seed schemes. Superfluorescence emission around 391 nm corresponding to the transition from \Bstate to \Xstate state is generated when a delayed seed laser is injected into the ionized nitrogen gas.
Ramsey fringes~\cite{Grynberg2010} show up in two following cases: 
(1) when another 391nm seed laser is injected and scanned over the relative delay $T_s+\tau_{ss}$ between the two seed pulses (see Fig.~\ref{fig1}(b));
(2) alternatively, shining another 800nm pump pulse and scanning over the temporal spacing $T_{p}+\tau_{pp}$ between the two pump pulses (see Fig.~\ref{fig1}(c)).
The major time delays $T_p=1.5$ps and $T_s=3$ps are introduced, in order to make the pairs of pump and seed pulses temporally separated, in this way, we can exclude the fringes formed by the direct interference of temporally overlapping pulse pairs.
The experimental observation of Ramsey interference fringes at 391nm emission intensity modulation as a function of $\tau_{pp}$ and with a period of 2.6fs corresponding to the transition frequency between the \Astate-\Xstate states\cite{ZhangAn2019Smos} suggests that \Astate and \Xstate are coherently populated by the pump laser, and it also provides a direct route to control the 391nm \Bstate-\Xstate emission by manipulating the \Astate-\Xstate coherence.

We simulate the N$_2^+$ air lasing process by employing the semiclassical optical-Bloch equation\cite{MacGillivray1976, Yuan2012, Weninger2014}. 
As demonstrated in Fig.~\ref{fig1}, a strong 800nm pump pulse propagates through a gas composed of nitrogen molecules, which does not only ionizes certain amount of neutral N$_2$ in the initial state $\vert 0\rangle$, but also coherently drives population transfer among five involved levels of N$_2^+$ cation, namely, {PPRM}. Via {PPRM} the population among N$_2^+$ levels is redistributed, and especially, significant amount of population is depleted from state $\vert1\rangle$ and transferred to high energy levels $\vert 2\rangle$, $\vert 3\rangle$ and $\vert 4\rangle$ in the $A^2\Pi_u$ manifold. 
When the population of state $\vert 5\rangle$ dominates over that of $\vert 1\rangle$ due to {PPRM},
the cloud of N$_2^+$ becomes a pencil-like active medium as in the experiment \cite{ZhangAn2019Cmos,ZhangAn2019Smos}. Subsequently, a 400nm deed pulse illuminates the population-inverted  N$_2^+$ gas and stimulates the emission from $\vert 5\rangle \rightarrow \vert 1\rangle$ transition.
The coupled optical-Bloch equation \cite{Arecchi1970, MacGillivray1976, Polder1979, Schuurmans1979, Yuan2012, Weninger2014} is used to describe our system.
\begin{eqnarray}
\partial_t \hat{\rho} &=& \frac{1}{i\hbar}\left[ \hat{H}, \hat{\rho}\right] + \hat{T} + \mathcal{L}\rho, \label{1}\\
\frac{1}{c} \partial_t E_p + \partial_z E_p &=& i \frac{2\pi n}{\epsilon_0} (\frac{P_{A}}{\lambda_{21}}\rho_{21} + \frac{P_{A}}{\lambda_{31}}\rho_{31} \nonumber \\ &+&  \frac{P_{A}}{\lambda_{41}}\rho_{41} + \frac{P_{B}}{\lambda_{51}} \rho^p_{51}),\label{eq2}\\
\frac{1}{c} \partial_t E_s + \partial_z E_s &=& i \frac{2\pi n}{\epsilon_0} \frac{P_{B}}{\lambda_{51}} \rho^s_{51},\label{eq3}
\end{eqnarray}
with initial and boundary conditions
\begin{eqnarray}
\rho_{ij}\left( 0, z\right) & = &\delta_{i0}\delta_{j0}, \\
E_p\left( 0, z\right) & = & E_s\left( 0, z\right) = 0, \\
E_p\left( t, 0\right) & = & E_{p1} \exp\left[ -\left( \frac{t-\tau_{p1}}{\sqrt{2}\tau}\right)^2 \right]\label{eq6}\\
& + &E_{p2} \exp\left[ -\left( \frac{t-\tau_{p2}}{\sqrt{2}\tau}\right)^2 +i \omega_p\left( T_p+\tau_{pp} \right) \right], \nonumber\\
E_{s}\left( t, 0\right) & = & E_{s1} \exp\left[ -\left( \frac{t-\tau_{s1}}{\sqrt{2}\tau}\right)^2 \right] \label{eq7}\\
& + &E_{s2} \exp\left[ -\left( \frac{t-\tau_{s2}}{\sqrt{2}\tau}\right)^2 +i \omega_s\left( T_s+\tau_{ss}\right)  \right],\nonumber
\end{eqnarray}
where $\delta_{ij}$ is the Kronecker delta symbol.
$\hat{\rho} $ is the density matrix of the five-level system in Fig.~\ref{fig1}(a), and $\hat{H}$ is the laser-molecular coupling Hamiltonian.
$\hat{T}=W_0 \rho_{00}\left( t \right) \exp\left[ -\frac{2\left( 2 I_p \right) ^{3/2}}{3 E_p\left( t \right) } \left( \frac{\sqrt{m}}{e\hbar}\right) \right] \sum_{i, j}F_{ij}\vert i\rangle\langle j\vert$ characterizes the tunneling ionization rate based on the ADK formular~\cite{ADK1987}.
$W_0$ the tunnelling ionization rate determined from the condition $\int_{-\infty}^{\infty}\hat{T}_{00}dt=0.9$ based on experimental observation \cite{Yao2016}.
$F_{ii}$ are the branching ratio of the \Np states and are defined as
\begin{equation}
F_{ii}^A=\left| \left\langle \phi_{vi}(A({\text N}_{2}^+))| \psi_0(X({\text N}_2)) \right\rangle  \right|^2 ,
\end{equation}
\begin{equation}
F_{ii}^B=\left| \left\langle \phi_{vi}(B({\text N}_{2}^+))| \psi_0(X({\text N}_2)) \right\rangle  \right|^2 .
\end{equation}
In order to determine $F_{ii}$, 
we solve Schr\"odinger equation based on discrete variable representation and ab initio potential~\cite{Yao2016}  
and obtain 
$F_{00}=-1$, $F_{11}=0.55825$, $F_{22}=0.05305$, $F_{33}=0.06259$, $F_{44}=0.05202$, $F_{55}=0.17885$.
(see details in Supplementary Material).
$E_{p\left( s \right) }$ is the slowly varying envelope of pump (seed) laser electric field.
$\mathcal{L}\rho$ effectively characterizes the spontaneous decay and other decoherence processes. Especially, $\mathcal{L}\rho_{41}=-\Gamma/2-\Gamma_a/2 $ and $\mathcal{L}\rho_{51}=-\Gamma/2-\Gamma_b/2 $.
$\rho_{41}$ and $\rho^{p\left( s\right) }_{51}$ represent the coherence of transition $\vert 1\rangle\rightarrow\vert 4\rangle$ and  $\vert 1\rangle\rightarrow\vert 5\rangle$ driven by pump (seed) laser respectively. The complete optical-Bloch equation and the simulation parameters are elaborated in the Supplementary Material.
%
\begin{figure}[t]
\vspace{-0.3cm}
\includegraphics[width=0.5\textwidth]{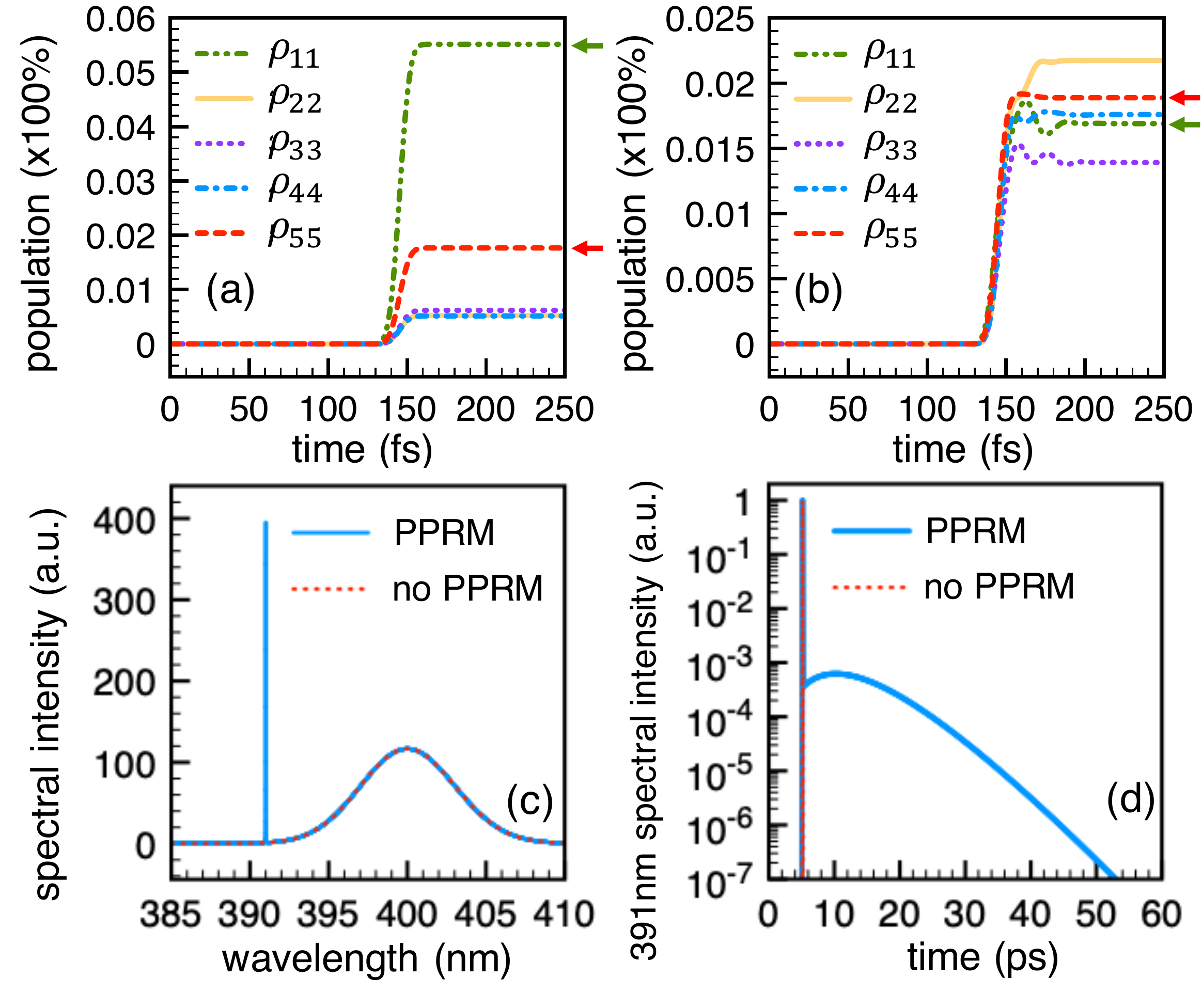}
\caption{\label{fig2}
(Color online) Population dynamics induced by pump pulse (a) without PPRM, (b) with PPRM. Green and red leftward arrows indicate $\rho_{11}$ and  $\rho_{55}$, respectively.
(c) The output spectra and (d) temporal evolution of the output spectral intensity at 391nm with (blue-solid line) and without (red-dotted line) PPRM.
}
\end{figure}
%
The emission spectrum is calculated by the Fourier transformation of the output seed pulse at $z=L$, namely,
\begin{equation}
S\left( \Delta \right) =\vert \int_{-\infty}^\infty E_s\left( t, L\right) e^{-i\Delta t}dt\vert^2,
\end{equation}
In order to prove the proposed {PPRM} mechanism, we demonstrate the pump-pulse induced dynamics of $N_{2}^{+}$ population without and with {PPRM} in Fig.~\ref{fig2}(a) and (b), respectively. 
{In Fig.~\ref{fig2}(a), all resulting population from mere ionization is given by the branching ratio $F_{ii}$, and especially there is no population inversion between state $\vert 5\rangle$ and $\vert 1\rangle$.}
However, when {PPRM} is included, the population inversion condition $\rho_{55}-\rho_{11}>0$ is significantly observed in the presence of the pump pulse (see red-dashed and green-dashed-dot lines in Fig.~\ref{fig2}(b)), which may play the key role to air lasing at 391nm.
{For this reason, in the output spectra shown in Fig.~\ref{fig2}(c), strong 391nm emission line emerges accompanying the PPRM process. Employing the continuous wavelet transform (CWT)~\cite{Lilly2008}, we extract the temporal evolution of the 391nm spectral intensity (see details in Supplementary Material). As illustrated in Fig.~\ref{fig2}(d), the 391nm emission under PPRM condition clearly demonstrates the dynamical establishment and decay of the macroscopic polarization on a 10-picosecond time scale.}
Our calculations reproduced previous experimental observations~\cite{Liu2015,ZhangAn2019Smos,ZhangAn2019Cmos}, in which the coherent 391nm emission is identified as superfluorescence.
To prove the population inversion is indispensable for the optical gain at 391nm, we simulate the process zero ionization rate of \Bstate (state $\vert 5\rangle$).
In Fig.~\ref{fig2a}(a), the temporal evolution of the seed laser shows modulation in form of a Bessel function~\cite{Crisp1970}.
Because the ionization rate of state $\vert 5\rangle$ is assumed to be zero, the population of state $\vert 5\rangle$ can just grow by PPRM.
From the above assumption, no population inversion is present, so the photons with 391nm wavelength will be absorbed (see Fig.~\ref{fig2a}(b)), namely, there is no optical gain but an absorption dip in the output spectra. 
%

\begin{figure}[t]
\vspace{-0.6cm}
\includegraphics[width=0.5\textwidth]{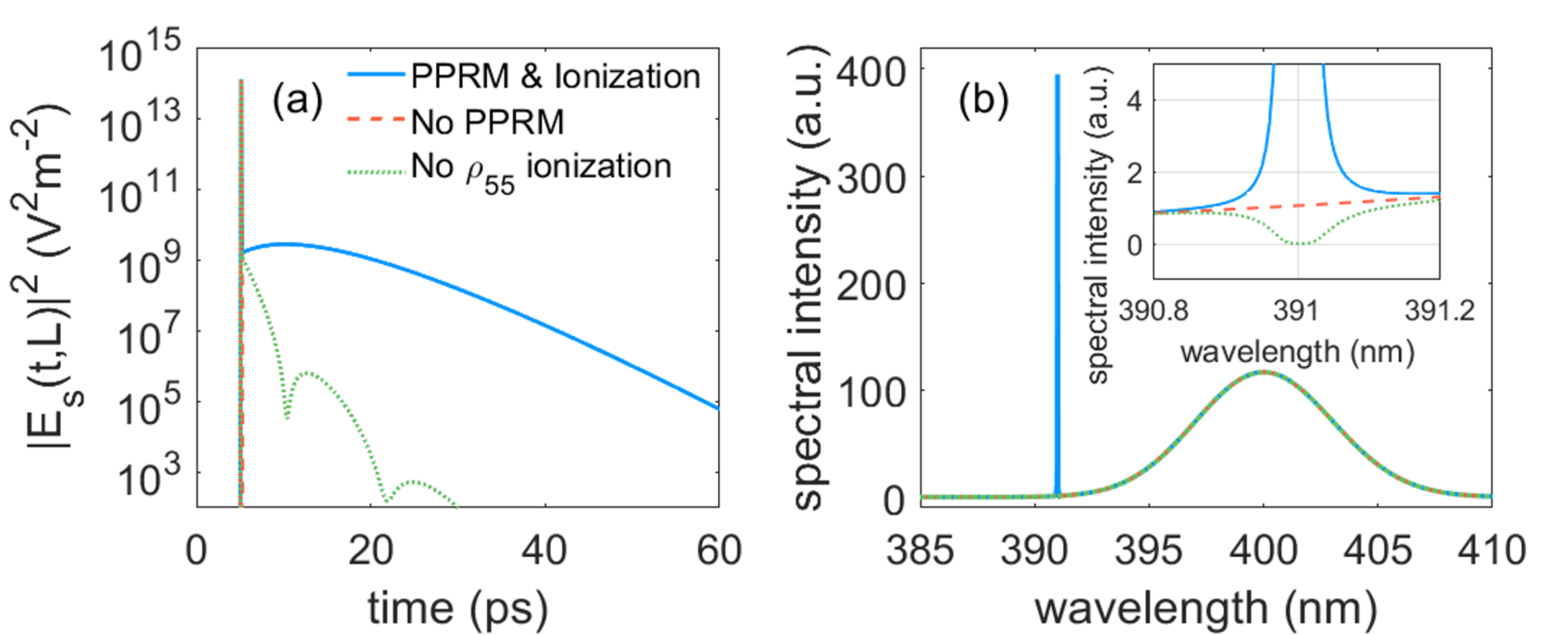}
\caption{\label{fig2a}
(Color online) Dynamics without population inversion. (a) Temporal evolution of the seed laser. (b) Output spectra with the assumption that no PPRM or no ionization  in \Bstate (setting $F_{55}=0$ for state $\vert 5\rangle$) is present.
 Two figures share the same legend.}
\end{figure}
	
\begin{figure}
\vspace{-0.3cm}
\includegraphics[width=0.5\textwidth]{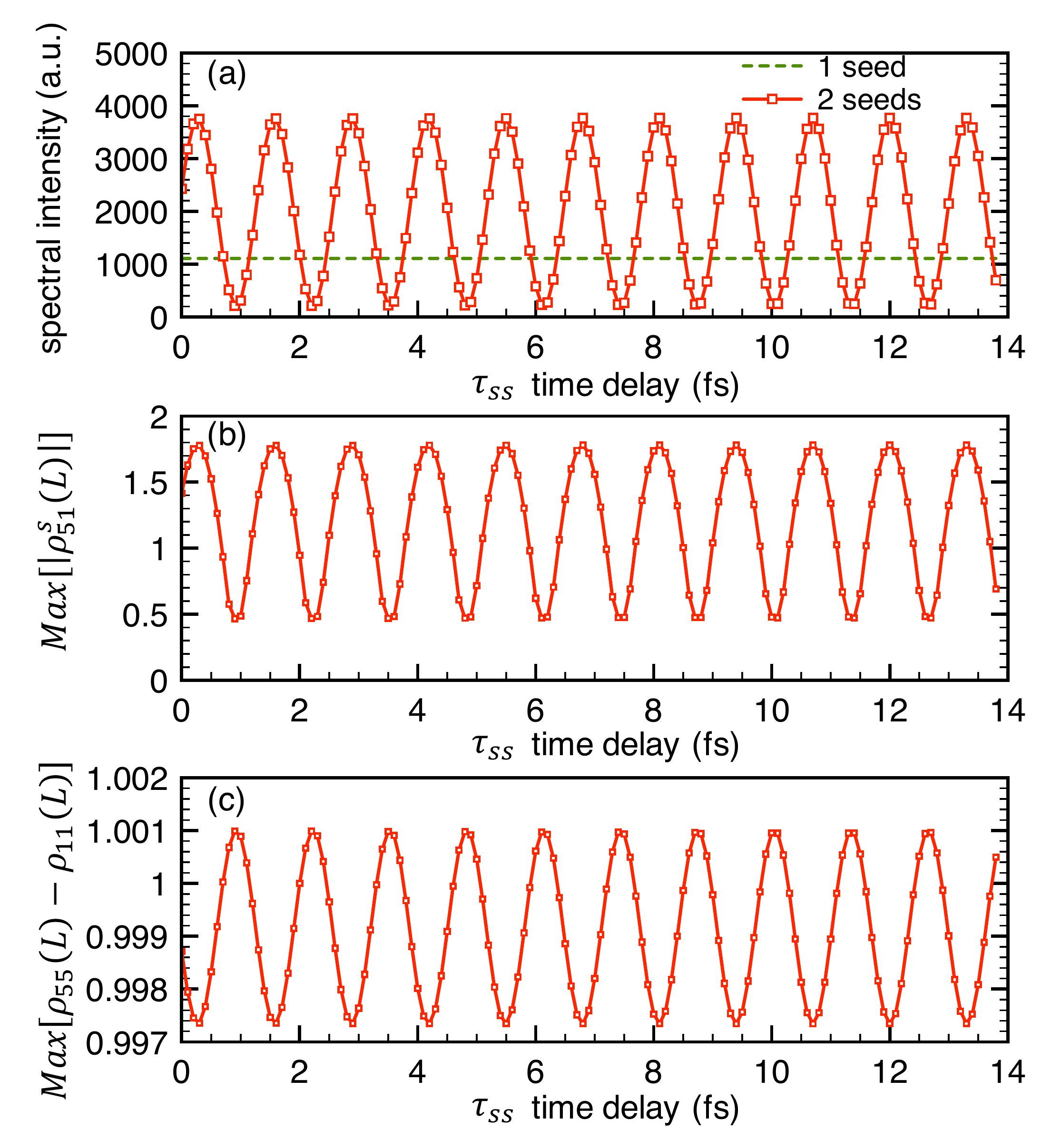}
\caption{\label{fig3}
(Color online) (a) $\tau_{ss}$-dependent spectral intensity at 391nm. Red solid line (green dashed line) is the intensity induced by two separated seed pulses (only one seed pulse). The major delay time between the two seed pulses is $T_s =3$ps and the pump-seed delay is $\tau_{sp}=0.1$ps.  The corresponding (b) maximum $\vert \rho_{51}^s\left( L\right) \vert$ and (c) maximum inversion $ \rho_{55}\left( L\right) -\rho_{11}\left( L\right) $, normalized to single-seed result, are demonstrated.
}
\end{figure}

In order to demonstrate the build-up dynamics of coherence and population inversion between state $\vert 5\rangle$ and $\vert 1\rangle$, we simulate the Ramsey fringes by subsequently illuminating the N$_2^+$ medium with two seed pulses spaced out adjustable time delay $T_s+\tau_{ss}$ apart, where the major delay $T_s=3$ps excludes the interference of overlapping light field. Fig.~\ref{fig3} shows the $\tau_{ss}$-dependent spectral intensity $S\left( \Delta \right)$, and the interference fringes clearly reveal the coherence between \Bstate and \Xstate states. 
And the modulation with 1.3fs period in Fig.~\ref{fig3} origins from the interference of the coherence of $\vert 1\rangle \rightarrow \vert 5\rangle $ transition pre-constructed  by the first seed pulse and the second seed pulse, which is consistent with observations \cite{ZhangAn2019Cmos}.
To further investigate the effect of double-seed scheme on N$_2^+$ dynamics, Fig.~\ref{fig3}(b) and Fig.~\ref{fig3}(c) illustrate the degrees of polarization and population inversion of the \Bstate and \Xstate states, characterized by the maximum density matrix elements $\rho_{51}$ and $\rho_{55}-\rho_{11}$ normalized to their corresponding single seed results. 
The Ramsey interference period of ~1.3fs corresponds to the transition frequency between \Xstate and \Bstate states, the interference fringes are in phase with the oscillation of polarization.
The degree of population inversion in Fig. \ref{fig3}(c) oscillates out of phase to the 391nm spectral intensity oscillation by half a period, as the stronger emission of 391nm photons weakens the degree of population inversion.
It is clear from Fig. \ref{fig3} that the amplitude of polarization oscillation is 2 orders of degree larger than that of the population inversion degree, this fact unambiguously demonstrates the dominance of coherence in altering the intensity of superfluorescence emission, despite of the indispensable requirement of population inversion between \Bstate and \Xstate states in the \Np system.
It paves the way to devise novel method for coherent control of the \Np air laser emission.

\begin{figure}
\vspace{-0.3cm}
\includegraphics[width=0.5\textwidth]{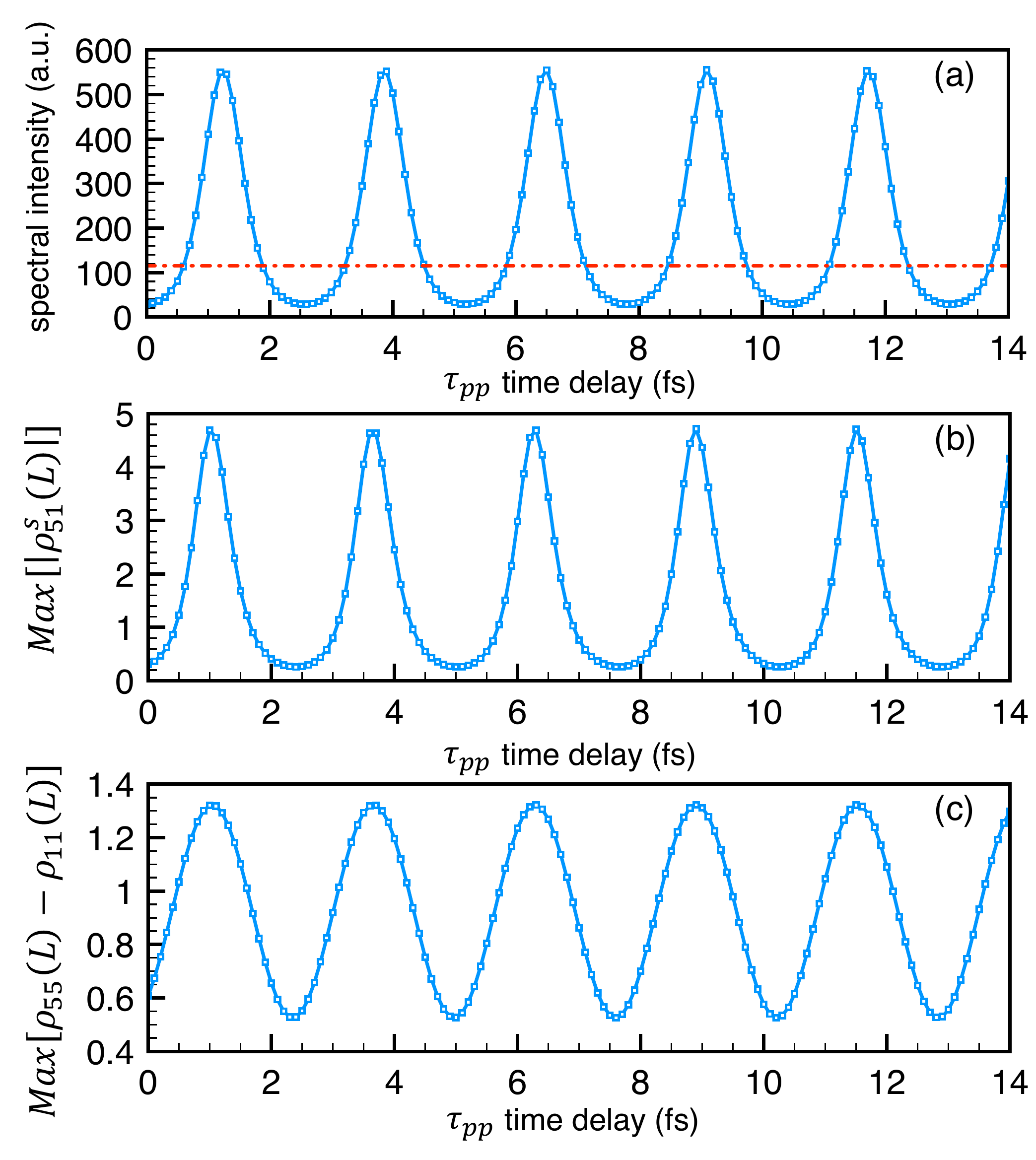}
\caption{\label{fig4}
(Color online)  (a) the spectral interference fringes at 391nm, (b) maximum $\vert\rho_{51}^s\left( L\right) \vert$ and, (c)  maximum $\rho_{55}\left( L\right)-\rho_{11}\left( L\right)$ are illustrated.
The latter two quantities are normalized to their single pump result.
The red dashed dotted line depicts the corresponding spectral intensity for cases with only one pump pulse.
}
\end{figure}

Fig.~\ref{fig4}(a) shows the 391nm signal intensity as a function of the time delay $T_p+\tau_{pp}$ between the two 800nm pump pulses with the major delay $T_p=1.5$ps, and $\tau_{sp}=5$ps, the second pump pulse is injected between the first pump and the seed pulses.
The period of modulation is about 2.6fs and corresponds to the transition frequency between state \Astate and state \Xstate.
We further demonstrate the effect of the second pump pulse on the coherence in Fig.~\ref{fig4}(b) as well as the population inversion of \Bstate-\Xstate state in Fig.~\ref{fig4}(c).
We normalize the polarization and population inversion degree between \Bstate and \Xstate with their values of the single-pump scheme.
For the Ramsey interference in the double pump scheme, the amplitude of oscillation of the population inversion degree becomes significantly larger and is on the same order of magnitude as the oscillation amplitude of the polarization Fig.\ref{fig4}(b)(c)). Besides, it is in phase with the intensity oscillation. This signature indicates that the population $\rho_{11}$ of the ground state \Xstate is strongly modulated by the second 800nm pump laser with the angular frequency of $2\pi c/800\text{nm}$.
It is a direct consequence from the fact that the \Astate and \Xstate must be coherently populated by the pump laser, and is consistent with coherent population transfer of the PPRM mechanism.
The double pump scheme can thus provide strong evidence for the coherent nature of the air laser, and exclusion of any description based on traditional laser rate equations.
Fig.\ref{fig4}(b) demonstrates strong dependence of 391nm emission intensity on the \Bstate-\Xstate polarization, which is induced by the 400nm seed pulse.

We have studied the coherence driven superfluorescence of the \Np air laser system, and have demonstrated the coherence between the \Astate-\Xstate as well as the \Bstate-\Astate pairs employing the Ramsey interferometry in the temporal domain. 
In the Ramsey interferometry, additional weak pulses of 800nm and 400nm wavelengths are injected and scanned over their relative time delays with respect to the major pump and seed pulses, which are resonant with the \Astate-\Xstate (800nm) and \Bstate-\Xstate (400nm) transitions.
The resulting Ramsey interference fringes, i.e. the 2.6fs and 1.3fs periodic oscillations in the intensity of 391nm emission from \Bstate to \Xstate transition as functions of relative delays between the two pump and the two seed pulses, have clearly illustrated the coherent population transfer and modulation of \Bstate-\Xstate polarization in the \Xstate-\AstateEl-\BstateEl system. 
Interestingly, although the double-seed scheme (Fig.~\ref{fig3}) and and double-pump scheme (Fig.~\ref{fig4}) can both enhance the 391nm emission, they are subject to different mechanisms, and thus provides variety of possibilities for the coherent control of the \Np air laser.
While lasing with two 391nm seed pulses, the \Bstate-\Xstate polarization has dominant contribution to the 391nm emission, the lasing with two 800nm pump pulses strongly relies on the degrees of both population inversion as well as polarization.
In the double-pump scheme, the second weak 800nm pump pulse tunes the \Astate-\Xstate coherence, and then indirectly affects the emission from \Bstate to \Xstate state transition, it enriches the understanding of the coherence relations of the \Xstate-\AstateEl-\BstateEl system as a complete picture, and shows that the superfluorescence emitted from \Bstate to \Xstate transition can be coherently controlled by the transfer between the \Astate and \Xstate state.
The consistency of the theory with forward lasing experiment\cite{ZhangAn2019Smos} sets the investigation of the backward lasing in the \Np system on a solid basis, which is suppressed by still unknown mechanisms and has not yet been experimentally realized.

Y.-H. K. and W.-T. L. are supported by the Ministry of Science and Technology, Taiwan (Grant No. MOST 107-2112-M-008-007-MY3 and Grant No. MOST 107-2745-M-007-001-). 
W.-T. L. is also supported by the National Center for Theoretical Sciences, Taiwan.
C. W. would like to acknowledge the support from the National
Natural Science Foundation of China (Grant Nos. 11625414 and 21673006).


\begin{widetext}

\appendix
\section{The optical-Bloch equations and simulation parameters}
In the following we present the complete optical-Bloch equations \cite{Scully2006} and detail of our simulation parameters for the \Np air laser system and the Ramsey fringes~\cite{Grynberg2010}.
The Hamiltonian reads $\hat{H} = \hat{H}_0 + \hat{H}_I$, where
\begin{eqnarray}
\hat{H}_0 &=& \hbar \omega_2 |2\rangle \langle2| + \hbar \left( \omega_2 + v_1 \right) |3\rangle \langle3| + \hbar \left( \omega_2 + v_2 \right) |4\rangle \langle4| + \hbar \omega_3 |5\rangle \langle5|,\\
\hat{H}_I &=& - \frac{\hbar}{2} \Omega_{p} e^{-i (\omega_p t - k_p z)} \left( |2\rangle \langle1| + |3\rangle \langle1| + |4\rangle \langle1| \right) - \frac{\hbar}{2} \left( \Omega_{f} e^{-i (\omega_p t - k_p z)} + \Omega_{s} e^{-i (\omega_s t - k_s z)} \right) |5\rangle \langle1| + H.c. .
\end{eqnarray}
Together with the transformations 
$\rho_{21} \rightarrow e^{-i (\omega_p t - k_p z)} \rho_{21}$, 
$\rho_{31} \rightarrow e^{-i (\omega_p t - k_p z)} \rho_{31}$,
$\rho_{41} \rightarrow e^{-i (\omega_p t - k_p z)} \rho_{41}$, 
$\rho_{51} \rightarrow e^{-i (\omega_s t - k_s z)} \rho_{51s} + e^{-i (\omega_p t - k_p z)} \rho_{51p}$, 
$\rho_{52} \rightarrow e^{-i (\omega_s t - \omega_p t - k_s z  +k_p z)} \rho_{52s} +  \rho_{52p}$,
$\rho_{53} \rightarrow e^{-i (\omega_s t - \omega_p t - k_s z  +k_p z)} \rho_{53s} + \rho_{53p}$, and
$\rho_{54} \rightarrow e^{-i (\omega_s t - \omega_p t - k_s z  +k_p z)} \rho_{54s} + \rho_{54p}$,
one can expand $\partial_t \hat{\rho} = \frac{1}{i\hbar}\left[ \hat{H}, \hat{\rho}\right] + \hat{T} + \mathcal{L}\rho$ and gets
\begin{eqnarray}
\frac{\partial \rho _{11}}{\partial t} &=& F_{11} w_0 \rho_0 \exp\left[{-\frac{2 (2 I_p)^{\frac{3}{2}} \sqrt{m}}{3 E_p q_e \hbar}}\right] + \Gamma_b \rho_{55} + \Gamma_a \rho_{44} + \Gamma_a \rho_{33} + \Gamma_a \rho_{22} \label{eqs1} \nonumber\\
& & + \frac{i}{2}[\Omega_p^* \rho_{21} + \Omega_p^* \rho_{31} + \Omega_p^* \rho_{41} + \Omega_f^* \rho_{51p} + \Omega_s^* \rho_{51s} - \rho_{51p}^* \Omega_f - \rho_{21}^* \Omega_p - \rho_{31}^* \Omega_p \nonumber\\
& & - \rho_{41}^* \Omega_p - \rho_{51s}^* \Omega_s] + e^{i(\omega_p-\omega_s)t}(\Omega_f^* \rho_{51s} - \rho_{51p}^* \Omega_s)
+ e^{-i(\omega_p-\omega_s)t}(\Omega_s^* \rho_{51p} - \rho_{51s}^* \Omega_f),\\
\frac{\partial \rho _{22}}{\partial t} &=& F_{22} w_0 \rho_0 \exp\left[{-\frac{2 (2 I_p)^{\frac{3}{2}} \sqrt{m}}{3 E_p q_e \hbar}}\right]  -\Gamma_a \rho_{22} + \frac{i}{2} (\rho_{21}^* \Omega_p-\rho_{21}\Omega_p^*),\\
\frac{\partial \rho _{33}}{\partial t} &=& F_{33} w_0 \rho_0 \exp\left[{-\frac{2 (2 I_p)^{\frac{3}{2}} \sqrt{m}}{3 E_p q_e \hbar}}\right] -\Gamma_a \rho_{33} + \frac{i}{2} (\rho_{31}^* \Omega_p-\rho_{31}\Omega_p^*),\\
\frac{\partial \rho _{44}}{\partial t} &=& F_{44} w_0 \rho_0 \exp\left[{-\frac{2 (2 I_p)^{\frac{3}{2}} \sqrt{m}}{3 E_p q_e \hbar}}\right] -\Gamma_a \rho_{44} + \frac{i}{2} (\rho_{41}^* \Omega_p-\rho_{41}\Omega_p^*),\\
\frac{\partial \rho _{55}}{\partial t} &=& F_{55} w_0 \rho_0 \exp\left[{-\frac{2 (2 I_p)^{\frac{3}{2}} \sqrt{m}}{3 E_p q_e \hbar}}\right] -\Gamma_b \rho_{55} \nonumber\\ 
& & - \frac{i}{2} \left[ \Omega_f^* \rho_{51p} + \Omega_s^* \rho_{51s}
- \rho_{51p}^* \Omega_f - \rho^*_{51s}\Omega_s + e^{i(\omega_p-\omega_s)t} (\Omega_f^* \rho_{51s} - \rho_{51p}^* \Omega_s) + e^{-i(\omega_p-\omega_s)t} (\Omega_s^* \rho_{51p} - \rho_{51s}^* \Omega_f)\right] ,\nonumber\\
\\
\frac{\partial \rho _{21}}{\partial t} &=& -\frac{\Gamma_a}{2} \rho_{21} \nonumber\\
& & -\frac{i}{2} \left[ {\Omega_p} \left(\rho_{32}^*+\rho_{42}^*-{\rho_{11}}+{\rho_{22}}\right)+\rho_{52p}^* \left({\Omega_f}+{\Omega_s} e^{i t ({\omega_p}-{\omega_s})}\right)\right.\left.+ {\rho_{52s}}^* \left({\Omega_s}+{\Omega_f} e^{-i t ({\omega_p}-{\omega_s})}\right)+2 {\rho_{21}} ({\omega_2}-{\omega_p})\right], \nonumber\\
\\
\frac{\partial \rho _{32}}{\partial t} &=& -\Gamma_a \rho_{32} -\frac{i}{2} \left(-{\Omega_p} \rho_{21}^*+{\rho_{31}} \Omega_p^* +2 {\rho_{32}} {v_ 1}\right),
\end{eqnarray}
\begin{eqnarray}
\frac{\partial \rho _{31}}{\partial t} &=& -\frac{\Gamma_a}{2} \rho_{31} \nonumber\\ 
& & -\frac{i}{2} \left[ {\Omega_p} \left(\rho_{43}^*-{\rho_{11}}+{\rho_{32}}+{\rho_{33}}\right)+\rho_{53p}^* \left({\Omega_f}+{\Omega_s} e^{i t ({\omega_p}-{\omega_s})}\right)\right.\left. + \rho_{53s}^* \left({\Omega_s}+{\Omega_f} e^{-i t ({\omega_p}-{\omega_s})}\right)+2 {\rho_{31}} ({v_ 1}+{\omega_2}-{\omega_p})\right],\nonumber\\
\\
\frac{\partial \rho _{41}}{\partial t} &=& -\left( \frac{\Gamma_a}{2} + \gamma_{41} \right) \rho_{41} \nonumber\\
& & - \frac{i}{2} \left[ \rho_{54p}^* \left({\Omega_f}+{\Omega_s} e^{i t ({\omega_p}-{\omega_s})}\right)+\rho_{54s}^* \left({\Omega_s}+{\Omega_f} e^{-i t ({\omega_p}-{\omega_s})}\right)+\Omega_p (-{\rho_{11}}+{\rho_{42}}+{\rho_{43}}+{\rho_{44}}) 
+ 2{\rho_{41}} ({v_ 2}+{\omega_2}-{\omega_p})\right],\nonumber\\
\\
\frac{\partial \rho _{42}}{\partial t} &=& -\Gamma_a \rho_{42} -\frac{i}{2} \left(2 v_2 \rho_{42} -{\Omega_p} \rho_{21}^*+{\rho_{41}} \Omega_p^*\right),\\
 \frac{\partial \rho _{43}}{\partial t} &=& -\Gamma_a \rho_{43} +\frac{i}{2} \left[ 2  \left( v_ 1 - v_ 2\right)  {\rho_{43}}+{\Omega_p} \rho_{31}^*-{\rho_{41}} \Omega_p^*\right] ,\\
\frac{\partial \rho _{51s}}{\partial t} &=& -\left( \frac{\Gamma_b}{2} + \Gamma \right) \rho_{51s} - \frac{i}{2} \left[ \Omega_s \left( {\rho_{55}}-{\rho_{11}}\right) +2 \left( \omega_3 - \omega_s \right)  \rho_{51s} +\Omega_p \left( \rho_{52s}+\rho_{53s}+\rho_{54s}\right) \right] ,\\
\frac{\partial \rho _{51p}}{\partial t} &=& -\left(  \frac{\Gamma_b}{2} + \Gamma \right)  \rho_{51p} -\frac{i}{2} \left[ {\Omega_f} ({\rho_{55}}-{\rho_{11}})+2 {\rho_{51p}} ({\omega_3}-{\omega_p})+{\rho_{52p}} {\Omega_p}+{\Omega_p} ({\rho_{53p}}+{\rho_{54p}})\right] ,\\
\frac{\partial \rho _{52s}}{\partial t} &=& -\left( \frac{\Gamma_a+\Gamma_b}{2} + \Gamma\right) \rho_{52s} +\frac{i}{2} \left[ 2  \left( {\omega_2}-{\omega_3}-{\omega_p}+{\omega_s}\right)  \rho_{52s}+  {\Omega_s} \rho _{21}^*-{\rho_{51s}} \Omega_p^* \right], \\
\frac{\partial \rho _{52p}}{\partial t} &=& -\left( \frac{\Gamma_a+\Gamma_b}{2} + \Gamma\right) \rho_{52p} -\frac{i}{2} \left[ 2 ({\omega_3}-{\omega_2})  \rho_{52p} -{\Omega_f} \rho _{21}^*+{\rho_{51p}} \Omega_p^*\right], \\
\frac{\partial \rho _{53s}}{\partial t} &=& -\left( \frac{\Gamma_a+\Gamma_b}{2} + \Gamma\right) \rho_{53s} +\frac{i}{2} \left[ 2 \left( {v_1}+{\omega_2}-{\omega_3}-{\omega_p}+{\omega_s}\right)  \rho_{53s}+ {\Omega_s} \rho_{31}^*-{\rho_{51s}} \Omega_p^* \right] ,\\
\frac{\partial \rho _{53p}}{\partial t} &=& -\left( \frac{\Gamma_a+\Gamma_b}{2} + \Gamma\right) \rho_{53p} +\frac{i}{2} \left[ 2  \left( {v_1}+{\omega_2}-{\omega_3}\right)  \rho_{53p}+{\Omega_f} \rho_{31}^*-{\rho_{51p}} \Omega_p^*\right], \\
\frac{\partial \rho _{54s}}{\partial t} &=& -\left( \frac{\Gamma_a+\Gamma_b}{2} + \Gamma\right) \rho_{54s} -\frac{i}{2} \left[ -{\Omega_s} \rho_{41}^*+{\rho_{51s}} \Omega_s^*-2 {\rho_{54s}} ({v_ 2}+{\omega_2}-{\omega_3}-{\omega_p}+{\omega_s})\right], \\
\frac{\partial \rho _{54p}}{\partial t} &=& -\left( \frac{\Gamma_a+\Gamma_b}{2} + \Gamma\right) \rho_{54p} +\frac{i}{2} \left[ 2 \left( {v_ 2}+{\omega_2}-{\omega_3}\right)  \rho_{54p} + {\Omega_f} \rho_{41}^*-{\rho_{51p}} \Omega_p^*\right], \label{eqs20}
\end{eqnarray}
The wave equations of pump and seed laser pulses read
\begin{equation}\label{eqs21}
\frac{1}{c} \partial_t E_p + \partial_z E_p = i \frac{2\pi n}{\epsilon_0} (\frac{P_{A}}{\lambda_{21}}\rho_{21} + \frac{P_{A}}{\lambda_{31}}\rho_{31} + \frac{P_{A}}{\lambda_{41}}\rho_{41} + \frac{P_{B}}{\lambda_{51}} \rho_{51p}),
\end{equation}
\begin{equation}\label{eqs22}
\frac{1}{c} \partial_t E_s + \partial_z E_s = i \frac{2\pi n}{\epsilon_0} \frac{P_{B}}{\lambda_{51}} \rho_{51s}.
\end{equation}
Together with initial and boundary conditions
\begin{eqnarray}
\rho_{ij}\left( 0, z\right) & = &\delta_{i0}\delta_{j0}, \\
E_p\left( 0, z\right) & = & 0, \\
E_s\left( 0, z\right) & = & 0, \\
E_p\left( t, 0\right) & = & E_{p1} \exp\left[ -\left( \frac{t-\tau_{p1}}{\sqrt{2}\tau}\right)^2 \right]\nonumber\\
& + &E_{p2} \exp\left[ -\left( \frac{t-\tau_{p2}}{\sqrt{2}\tau}\right)^2 +i \omega_p\tau_{pp} \right],\\
E_{s}\left( t, 0\right) & = & E_{s1} \exp\left[ -\left( \frac{t-\tau_{s1}}{\sqrt{2}\tau}\right)^2 \right] \nonumber\\
& + &E_{s2} \exp\left[ -\left( \frac{t-\tau_{s2}}{\sqrt{2}\tau}\right)^2 +i \omega_s\tau_{ss} \right],
\end{eqnarray}
Because the N$_2$ molecule is not pre-aligned in space, and the multiphoton ionization does not necessarily produce the $\cos^2\theta$ distribution along the polarization axis of the ionization laser, given the fact that the outgoing electron is not only in the $s$-wave state, we can thus assume uniform distribution of the \Np cation in the angular direction. Given the pump and seed pulses have the same linear polarization, the dominant contribution comes from the \Np cation with $\pi/4$ angle between the molecular axis and the laser polarization, and the pump and seed pulses have equal contribution to the vertical \Xstate-\AstateEl and parallel \Xstate-\Bstate transitions following their selection rules.
The numerical methods used for the dynamics of density matrix, namely, Eq.~(\ref{eqs1}-\ref{eqs20}) and the wave equation, i.e., Eq.~(\ref{eqs21} \& \ref{eqs22}), are 4th order Runge-Kutta and Lax method, respectively. 
%
Our one-dimensional numerical calculations for the pencil-like N$_2$ sample are performed in a simulation box $\left[ 0,L\right]$, where laser pulses enter the medium at $z=0$ and exit at $z=L$. 
The pulse propagation effects crucially depend on optical depth $n\sigma L$, where $n$ is the particle number density, $\sigma$ the resonant absorption cross section and $L$ the medium length. 
For a given optical depth and a $\sigma$, no significant difference is observed when varying $n$ and $L$ in the numerical calculation. One thus has flexible choices of $n$ and $L$ for a given $nL=$ constant to ease the lengthy computation time.

The seed spectra are calculated by
\begin{equation}
S\left( \Delta \right) =\vert \int_{-\infty}^\infty E_s\left( t, L\right) e^{-i\Delta t}dt\vert^2,
\end{equation}
namely, the Fourier transformation of the output seed pulses.
Here $\Delta = 108.47$ rad$\cdot$THz represents the spectral intensity at 391nm.
Obtaining the electric field $E_s(t,L)$ of the output seed pulse, we then employ the continuous wavelet transform to analyze the temporal evolution of the key 391nm emission intensity at the lasing wavelength.
We use the generalized Morse wavelets with symmetry parameter to be 3 and time-bandwidth product to be 60~\cite{Lilly2008}.
The continuous wavelet analysis of the the output seed pulse is presented in Fig.~\ref{figS1}.
\begin{figure}
    \centering
    \includegraphics[width=0.75\textwidth]{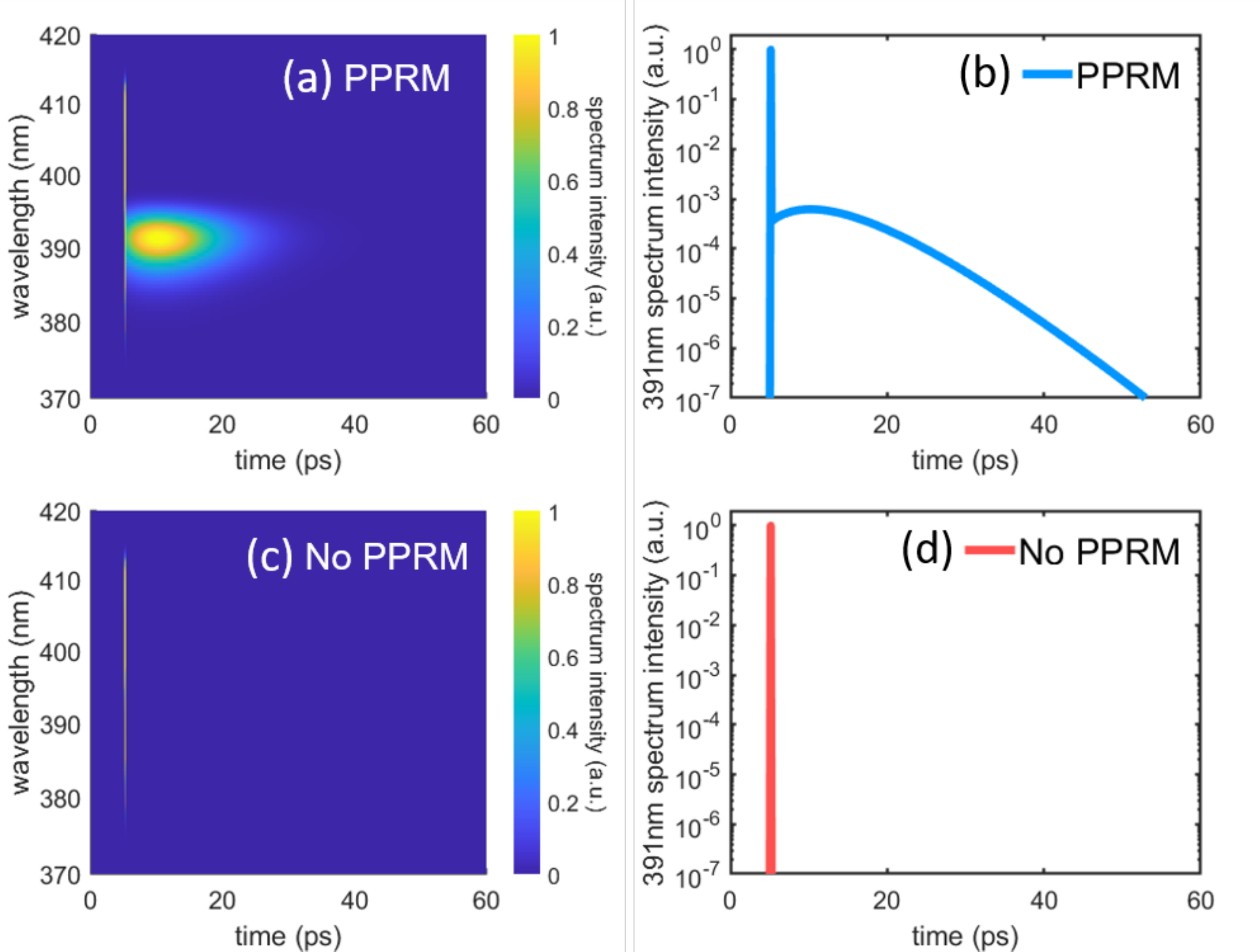}
    \caption{(a) The continuous wavelet transform (CWT) of the output seed pulse $E_s(t, L)$ with PPRM. (b) The temporal evolution of the 391nm spectral intensity with PPRM. (c) The CWT of the output seed pulse $E_s(t, L)$ without PPRM. (d) The temporal evolution of the 391nm spectral intensity without PPRM.}
    \label{figS1}
\end{figure}
\begin{table}[h]
\centering
\caption{\label{table1}
Notations used throughout the text.
}
\begin{tabular}{l c l}
\hline
\hline
\multicolumn{1}{c} {Notation (unit)} & {Value} & {Explanation}\\
\hline
$c\left( \mathrm{m/s}\right)$                   & $3\times 10^8$ & the speed of light in vacuum\\
$\epsilon_0\left( \mathrm{F/m}\right)$                             & $8.85\times 10^{-12}$ &  vacuum permittivity \\
$\hbar\left( \mathrm{m^2 kg/s}\right)$   & $1.054\times 10^{-34}$ & reduced Planck constant \\
$e\left( \mathrm{C}\right)$                      & $1.602\times 10^{-19}$ & electronic charge\\
$m\left( \mathrm{kg}\right)$                   & $9.11\times 10^{-31}$ & electronic mass\\
$A_p\left( \mathrm{W/cm^2}\right) $        & $2.2 \times 10^{14}$ & peak intensity of pump pulse 1\\
$A_{p2}\left( \mathrm{W/cm^2}\right) $       & $9.9 \times 10^{11}$ (Fig.~4)& peak intensity of pump pulse 2\\
$A_s\left( \mathrm{W/cm^2}\right) $        & $9.259 \times 10^{3}$ (Fig.~3) &    peak intensity of seed pulse 1\\
                                                                & $5.556 \times 10^{4}$ (Fig.~4) &  \\
$A_{s2}\left( \mathrm{W/cm^2}\right) $      & $9.259 \times 10^{3}$ (Fig.~3) &    peak intensity of seed pulse 2\\
$P_A\left(  \mathrm{C\cdot m} \right) $             & $2.21 \times 10^{-30}$ & dipole moment for \\
&  & transition $X^2\Sigma_g^+\rightarrow A^2\Pi_u$\\
$P_B\left(  \mathrm{C\cdot m} \right) $             & $6.41 \times 10^{-30}$ & dipole moment for \\ 
&  & transition $X^2\Sigma_g^+\rightarrow B^2\Sigma_u^+$\\
$I_p\left(  \mathrm{eV} \right)$                & $15.58$ & internal binding energy \\
$n\left(  \mathrm{mm^{-3}} \right)$                                            & $1.237 \times 10^{17}$ & particle density of N$_2$ gas\\
$L\left( \mathrm{mm}\right) $                & 0.01 & medium length\\
$\Gamma\left( \mathrm{THz}\right) $     & $0.2$ & collision rate\\
$\Gamma_a\left( \mathrm{THz}\right) $  & $2\times 10^{-5}$ & decay rate of state $A^2\Pi_u$\\
$\Gamma_b\left( \mathrm{THz}\right) $  & $1.017\times 10^{-7}$ & decay rate of state $B^2\Sigma_u^+$\\
$ \tau_{p1} \left( \mathrm{fs}\right)$         & $113$ & peak time of pump pulse 1\\
$ \tau_{pp} \left( \mathrm{ps}\right)$        & $-4 - 25$ (Fig.~4) & delay time $ \tau_{p2} - \tau_{p1} $\\
$ \tau_{p2} \left( \mathrm{ps}\right)$        & $\tau_{p1}+\tau_{pp}$ (Fig.~4) & peak time of pump pulse 2\\
$ \tau_{s1} \left( \mathrm{fs}\right)$        & $213$ (Fig.~3) & peak time of seed pulse 1\\
                                                                  & $5113$ (Fig.~4) &  \\
$ \tau_{s2} \left( \mathrm{fs}\right)$        & $213-227$ (Fig.~3) & peak time of seed pulse 2\\
$\tau_{sp}\left( \mathrm{ps}\right)  $	      & $0.1$  (Fig.~3)         & delay time $ \tau_{s1} - \tau_{p1} $ \\
                                                                 & $5$  (Fig.~4)            &\\
$\tau_{ss}\left( \mathrm{fs}\right)  $	      & $0-14$ (Fig.~3) & delay time $ \tau_{s2} - \tau_{s1} $\\
$\lambda_p\left( \mathrm{nm}\right)  $    & $800$ &central wavelength of pump pulses \\
$\lambda_s\left( \mathrm{nm}\right)  $    & $400$ &central wavelength of seed pulses\\
$\lambda_{21}\left( \mathrm{nm}\right)  $ & $1109$ &wavelength of transition $\vert 1 \rangle\rightarrow\vert 2 \rangle$\\
$\lambda_{31}\left( \mathrm{nm}\right)  $ & $919$ &wavelength of transition $\vert 1 \rangle\rightarrow\vert 3 \rangle$\\
$\lambda_{41}\left( \mathrm{nm}\right)  $ & $786.5$ &wavelength of transition $\vert 1 \rangle\rightarrow\vert 4 \rangle$\\
$\lambda_{51}\left( \mathrm{nm}\right)  $ & $391$ &wavelength of transition $\vert 1 \rangle\rightarrow\vert 5 \rangle$\\
$\omega_p\left( \mathrm{rad/s}\right)  $  & $2.36\times 10^{15}$ &angular frequency of   seed laser\\
$\omega_s\left( \mathrm{rad/s}\right)  $  & $4.82\times 10^{15}$ &angular frequency of  pump laser\\
$\Delta_2\left( \mathrm{rad/s}\right) $     &  $6.67\times 10^{14}$  & detuning $\omega_p - 2\pi c/\lambda_{21}$\\
$\Delta_3\left( \mathrm{rad/s}\right) $     &  $3.05\times 10^{14}$ & detuning $\omega_p - 2\pi c/\lambda_{31}$\\
$\Delta_4\left( \mathrm{rad/s}\right) $     &  $-4.04\times 10^{13}$ & detuning $\omega_p - 2\pi c/\lambda_{41}$\\
$\Delta^p_5\left( \mathrm{rad/s}\right) $   &  $-2.46\times 10^{15}$ & detuning $\omega_p - 2\pi c/\lambda_{51}$\\
$\Delta^s_5\left( \mathrm{rad/s}\right) $   & $0$ & detuning $\omega_s - 2\pi c/\lambda_{51}$\\
$\tau\left( \mathrm{fs}\right)  $           & $20 $ & pulse duration of laser pulse\\
$W_0\left( \mathrm{Hz}\right)$              & $1.78927 \times 10^{19} $ & tunnelling ionization rate \\
$\rho_{ii}  $      & based on simulation  & population of state $\vert i \rangle$\\
$\rho_{ij}  $      & based on simulation  & coherence between state $\vert i \rangle$ and $\vert j \rangle$ (i $\neq$ j)\\
$E_{p(s)}\left( \mathrm{V/m}\right)  $      & based on simulation  & electric field of pump (seed) laser\\
$\Omega_{p(s)}\left( \mathrm{rad/s}\right)  $      & $P_{A(B)} E_{p(s)}/\hbar $ & Rabi frequency of pump(seed) laser\\
$\Omega_{f}\left( \mathrm{rad/s}\right)  $      & $P_B E_p/\hbar $ & Rabi frequency of pump laser driving transition $\vert 1 \rangle\rightarrow\vert 5 \rangle$ \\
$\omega_2\left( \mathrm{rad/s}\right)  $    & $1.7 \times 10^{15} $ & $\vert 1 \rangle\rightarrow\vert 2 \rangle$ transition angular frequency\\
$\omega_3\left( \mathrm{rad/s}\right)  $    & $4.82 \times 10^{15} $ & $\vert 1 \rangle\rightarrow\vert 5 \rangle$ transition angular frequency\\
$v_1\left( \mathrm{rad/s}\right)  $         & $3.51 \times 10^{14} $ & $\vert 2 \rangle\rightarrow\vert 3 \rangle$ transition angular frequency\\
$v_2\left( \mathrm{rad/s}\right)  $         & $6.97 \times 10^{14} $ & $\vert 2 \rangle\rightarrow\vert 4 \rangle$ transition angular frequency\\
\hline
\hline
\end{tabular}
\end{table}

\clearpage

\section{The determination of branching ratio $F_{ii}$}

We use the ab initio potential of neutral N$_2$ and N$_2^+$ cation \cite{Yao2016} to solve the vibrational Schr\"odinger equation in the discrete variable representation (DVR)\cite{Colbert1992}. The potential curves of the electronic ground state $X^1\Sigma_g^+$ of N$_2$ molecule and the \Xstate, \AstateEl and \Bstate states of \Np cation are shown in Fig.~\ref{figS2}.
In the discrete variable representation, the Hamiltonian matrix elements are written as\cite{Colbert1992}
\begin{eqnarray}
	H_{ii'}= \left\{
\begin{aligned}
&	\frac{(-1)^{i-i'}}{2\Delta x^2} \frac{\pi^2}{3}+V(x_i)\,, &\quad i=i' \\
&	\frac{(-1)^{i-i'}}{2\Delta x^2} \frac{2}{(i-i')}\,, &\quad i \neq i'
\end{aligned}
\right.
\end{eqnarray}
We then solve the Schr\"odinger's equation to get the branching ratios $F_{ii}$, which are defined as
\begin{eqnarray}
F_{ii}^A&=&\left| \left\langle \phi_{vi}(A(\text{N}_{2}^+))| \psi_0(X(\text{N}_2)) \right\rangle  \right|^2 \nonumber\\
F_{ii}^B&=&\left| \left\langle \phi_{vi}(B(\text{N}_{2}^+))| \psi_0(X(\text{N}_2)) \right\rangle  \right|^2 
\,.
\end{eqnarray}
%

\begin{figure}[htb]
	\vspace{-0.5cm}
	\includegraphics[width=0.5\textwidth]{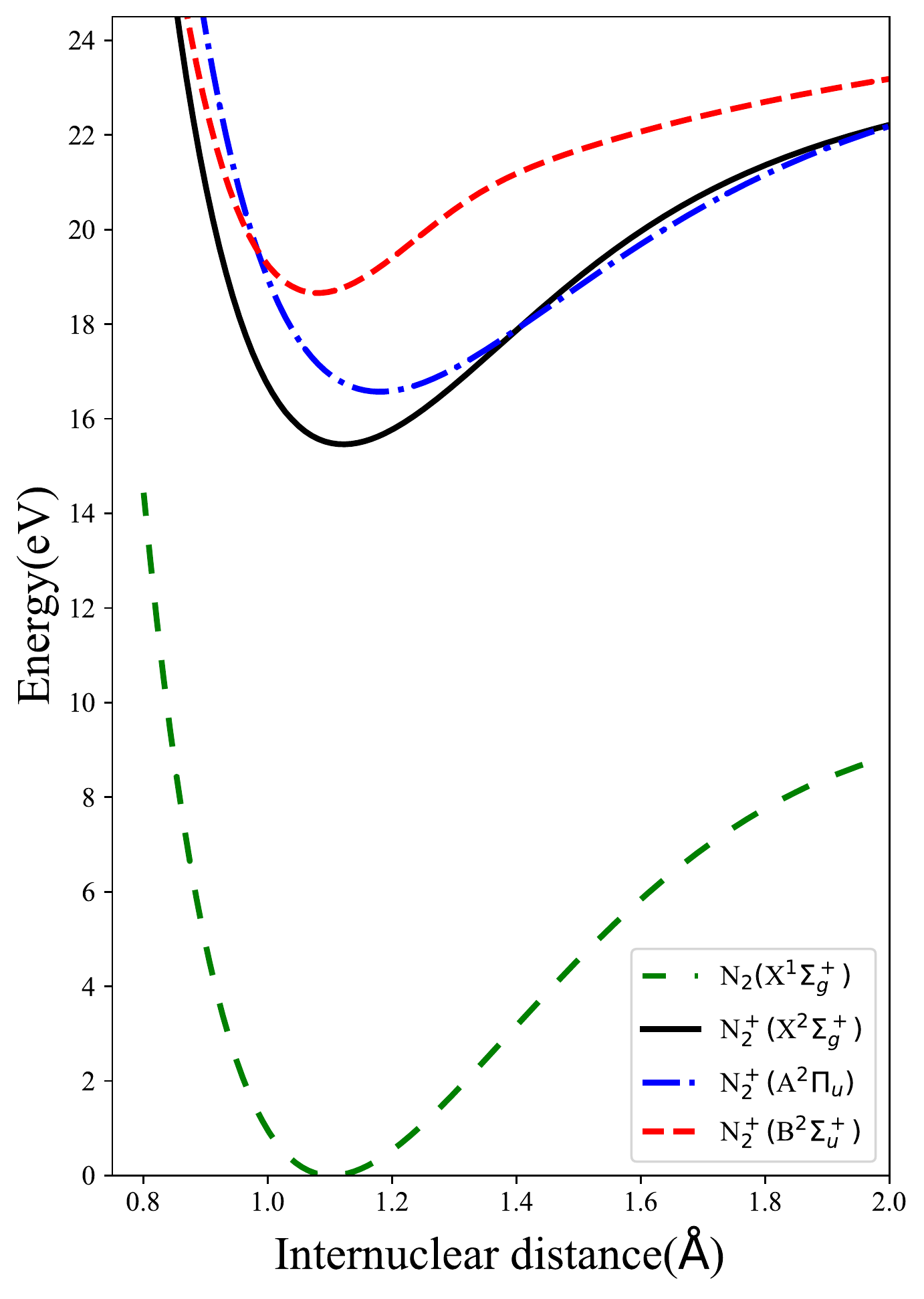}
	\caption{\label{figS2}
		Potential energy curves of the electronic ground state $X^1\Sigma_g^+$ of N$_2$ molecule and the \Xstate, \AstateEl and \Bstate states of \Np cation.\cite{Yao2016}.
	}
\end{figure}

\end{widetext}
\clearpage

\bibliographystyle{apsrev}
\bibliography{20181002SF}

\end{document}